# Observation of multi-soliton asynchronous buildup dynamics in all-PM mode-locked lasers

XUEMING LIU,* XIAOXIANG HAN, AND YUSHENG ZHANG

*State Key Laboratory of Modern Optical Instrumentation, College of Optical Science and Engineering, Zhejiang University, Hangzhou 310027, China*
*Corresponding author: liuxm@zju.edu.cn*

**Ultrafast transient phenomena are the most important research field in nonlinear systems, and the mode-locking fiber laser provides an excellent research platform for ultrafast transient phenomena. We report on the experimental observation of multi-soliton asynchronous buildup dynamics in all-polarization-maintaining mode-locked fiber laser. The build-up dynamics display several novel features. The different solitons in multi-soliton generation are produced asynchronously instead of simultaneously. Solitons generated at different time have different propagation speeds at metastable state and same speeds at stable state, indicating that different solitons have different energy at metastable state. The intra-cavity energy exhibits quantization property with respect to the build-up process. A proportion of background pulse vanished with respect to the birth of new pulse, which shows the jungle law in fiber laser system. The buildup dynamics may benefit the research of ultrafast pulse generation, interaction, regulation, energy improvement, and stability improvement.**

The ultrafast phenomena have attracted a great deal of research interest, especially the transient, non-repetitive events [1]-[3]. Ultrafast laser is an excellent platform to study the ultrafast phenomena [4]-[7]. Fiber laser have a wide range of application, including fiber optical communications, industry shipbuilding, automobile manufacturing, military defense security, medical equipment and so on. The research task of real-time buildup dynamics in fiber laser is extremely urgent due to its rich variety of highly stochastic and non-repetitive phenomena, such as soliton rain [8], harmonic mode-locking [9], soliton collision [10], and multi-wavelength mode-locking [11]. The behavior of multi-pulse cluster has been mostly studied by numerical simulations and temporal observation. There are a variety of theories to explain the formation of solitons. Modulation instability is closely related to the formation of soliton in fiber laser, which enables the noise on the CW or quasi-CW to grow exponentially [12][13]. When the pump power exceeds certain threshold, multi-pulse generation is a general behavior in laser system [14]-[16]. The increasing pump power leads to the growth of soliton peak power, and then the high peak power would result in that the soliton transmission becomes smaller than the linear cavity transmission. So when the pump power is increased, the soliton is less amplified than that of the CW background. Then the background pulse with a certain frequency satisfies the lasing condition may grow up into a new soliton, which leads to multi-soliton operation. Kurtner *et al.* have reported that multi-soliton operation is caused by pulse splitting [17]. Lederer *et al.* have reported that multi-soliton phenomenon is caused by band-pass filter effect based on Ginzburg Landau equations [18]. When multi-soliton are in stable state, each soliton exhibits same parameters, which is caused by energy quantization effect [19][20]. Energy quantization effect limits the energy of single pulse. In order to create a robust experimental platform, all-polarization-maintaining (all-PM) fiber lasers are proposed, such as a figure-eight laser, and an Er-doped laser using carbon nanotube saturable absorber [21]-[23]. Since the linear polarization state is always maintained in those all-PM fiber lasers, all-PM fiber lasers have excellent long-term stability, which can be used as excellent experimental platform for investigating pulse formation dynamics in laser cavities. Although both simulation and experiment have simulated and produced multi-pulse under the appropriate cavity parameters, the real-time observation of multi-soliton asynchronous buildup dynamics have not been reported, which requires ultrafast analyzing technique.

There are lots of transient, stochastic phenomena in our surrounding environment, such as chemical reaction, phase transformation [24]. The dispersion Fourier transform (DFT) is a very effective method of high speed measurement [25]-[27]. When a soliton is propagated in the dispersive medium, it is stretched due to the group velocity dispersion (GVD), while the different wavelengths in the spectrum are separated in temporal domain due to the difference of propagation velocity. The shape of soliton obtained after stretching in DCF is in accordance with the shape of soliton spectrum, thus realizing spectrum detection of target soliton in temporal domain [28]-[31]. The high speed analog digital converter (ADC) sampling can be used to analyze the pulse spectrum in temporal domain, and the detection rate is much higher than that of the traditional spectrometer [32][33]. The DFT technique opens a portal to directly observe the ultrafast dynamics of soliton generation, the interaction of soliton molecules [34][35], and soliton explosions [36]-[38]. Erkintalo *et al.* and Liu *et al.* have reported the measurements of soliton explosions by using the TS-DFT technique [39][40].

In this paper, we have demonstrated the real-time observation of multi-soliton asynchronous buildup dynamics in passively mode-locked erbium-doped fiber (EDF) laser with a carbon nanotubes (CNT)-polyvinyl alcohol (PVA) mode-locker. During the buildup process, we have observed several unique features. Each soliton of multi-soliton generated at different round-trips, and the time interval varies from thousands to ten thousands round-trips, which indicating the asynchronous process of mode-locking. Solitons generated at different time exhibit different pulse energy at metastable state and same energy at stable state, which shows the energy distribution with quantization property. A proportion of background pulse vanished with respect to the birth of new soliton, which shows the law of jungle in nonlinear system. The intensity variation confirms that there is energy exchange between old soliton and resonator system. The Q-switched lasing intensity of new soliton is much lower than that of born-early soliton, which indicating that the degree of mode-locked difficulty became smaller than that of previous soliton. The buildup dynamics may benefit the research of ultrafast pulse generation, interaction, regulation, energy improvement, background pulse suppression, and stability improvement. The unique features may shed some light on the internal physics law of ultrafast system.

The setup of the proposed fiber laser is shown in Fig. 1. In the fiber laser cavity, 1-m PM-EDF with 110 dB/m peak absorption at 1530 nm acts as the gain media, and the rest fibers are single-mode, polarization maintaining (PANDA style) fiber. The mode-locker is assembled by sandwiching a ~2 mm$^2$ free-standing CNT-PVA composite film between two fiber ferrules inside a physical contact ferrule connector. A 980 nm laser diode (LD) provides pump via the wavelength-division multiplexer (WDM) port of the hybrid integrated device (i.e. PM-WTI). The hybrid device is integrated with WDM, 10/90 tap coupler and isolator. The hybrid device has greatly simplified the cavity structure. The LD was controlled by a fast optical chopper system, which can turn on and off of the pump source. All fibers and components in the proposed laser were spliced by using the fusion splicer with $\theta$-axis rotation (Fujikura FSM-100P+) to match the appropriate polarization axis. An autocorrelator (AC), an optical spectrum analyzer (OSA) and a radio-frequency (RF) analyzer are employed to monitor the stable state of output soliton simultaneously. The buildup dynamics of output soliton is monitored by real-time oscilloscope via DFT process. The output is splitted into two pulses with a 50:50 coupler. One of the separated pulse inject into the real-time oscilloscope (20 GSa/s sampling rate) via a high-speed photodetector (12 GHz bandwidth) directly, and the other of the separated pulse passes through a section of dispersion-compensating fiber (DCF) before detection.

The continuous wave (CW) is generated at the pump power of 20 mW. Then we swiftly increase the pump power to 40 mW, multiple solitons are formed in the laser cavity. The ordinary spectrum obtained by OSA and real-time spectral acquisition are measured simultaneously. Comparing to the ordinary spectrum, the real-time spectral data is delayed in time by ~24.5 μs due to the extra transmission in 5 km DCF. Figure 2 shows the 3D buildup process of multiple solitons, ordinary spectrum directly obtained by OSA and AC trace. The specific mapping relation between temporal and spectral domain can be given as $\Delta t=|D|L\Delta\lambda$, where $\Delta\lambda$ is the bandwidth of optical spectrum, $\Delta t$ is the corresponding time spacing, $D$ and $L$ are the dispersion parameter and length of DCF, respectively. The projection shows the temporal evolution along with the round-trips. The experimental observations reveal that there exists an evidently asynchronous birth course before the appearance of stable soliton trains (i.e., stable mode-locking). As we can see that the peak of Q-switched lasing for each soliton has different values and become smaller with respect to roundtrips, which confirms that the mode-locking of later soliton is become easier than that of before. The Q-switched lasing peak power of the 4$^{th}$ soliton is much lower than that of 1$^{st}$, 2$^{nd}$ and 3$^{rd}$, which means that the mode-locked environment of resonant have substantial improvement. So we can safely deduce that the degree of mode-locked difficulty became smaller with respect to the passing round-trips.

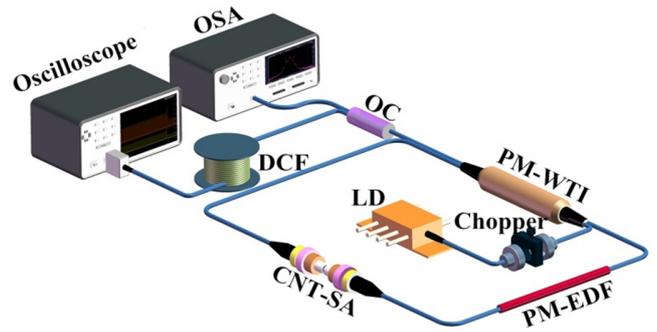

Fig. 1. Experimental setup of the all-PM fiber laser. LD, laser diode; PM-WTI, polarization maintaining hybrid integrated device of wavelength division multiplexer, tap coupler and isolator; PM-EDF, polarization maintaining erbium-doped fiber; PD, photodetector; CNT-SA, carbon nanotubes saturable absorber; OSA, optical spectrum analyzer; OC, optical coupler; DCF, dispersion compensating fiber.

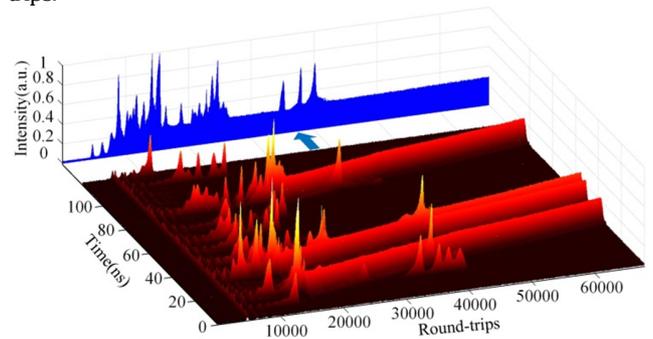

Fig. 2. Experimental 3D real-time characterization of multiple soliton. The intensity profile evolves along with the round-trip time and the round-trip number. The projection shows the temporal evolution along with the round-trips.

The complex formation process of multi-soliton reveals several unique features. As we can see, the background pulses still exist after the first soliton is born, and the background pulses are completely vanished until the last stable soliton is generated. Figure 3 depicts the experimental observation of multi-soliton 2D build-up dynamics. It can be seen that there are multiple relaxation oscillations before the intense interaction between pulses. The relaxation oscillation can induce the generation of Q-switched lasing, which includes multiple subordinate pulses together with a dominant pulse competing for energy to form soliton. After fierce competition, beating process with organized interference fringes begins, which shows the interaction between the remaining dominant and only one subordinate pulse. The expanded views of beating process from first and last solitons are shown in Fig. 3(b)-(c), respectively. The birth times of soliton 1 2, 3, and 4 are 14800RT, 16375RT, 25350RT, 41250RT, respectively. The time interval is different and become larger with respect to the passing round-trips. As nonlinearity and anomalous dispersion are still present, the random background pulse can effectively seed modulation instability, which can generate multiple solitons. Each beating process is corresponding to a new-born soliton, while the processes of background pulse fail to achieve mode-locking does not have beating process (i.e. those background pulses eventually cease to exist). The strong fluctuations before broadband mode-locking and spectral oscillation are the signature of self-phase modulation (SPM) effect, which plays a dominant role in mode-locking operation. After this initial SPM spectral broadening, the dominant pulse is rapidly evolved into a soliton with smooth and broad spectrum. When the 3rd soliton is generated, the background pulses during the region of 40-70 ns vanished as shown in Fig. 3(a). Similarly, during the 4th pulse formation, background pulses are disappeared in the region of 10-20 ns. The soliton circulating in the cavity will periodically reshape itself and shed the excess energy to background pulse so that it can reshape itself to a stable state [31]. The resonant energy transfer occurs between dispersive waves and background pulse, whose phase is matched to the soliton, and then the matched background pulse will grow into a new-born soliton.

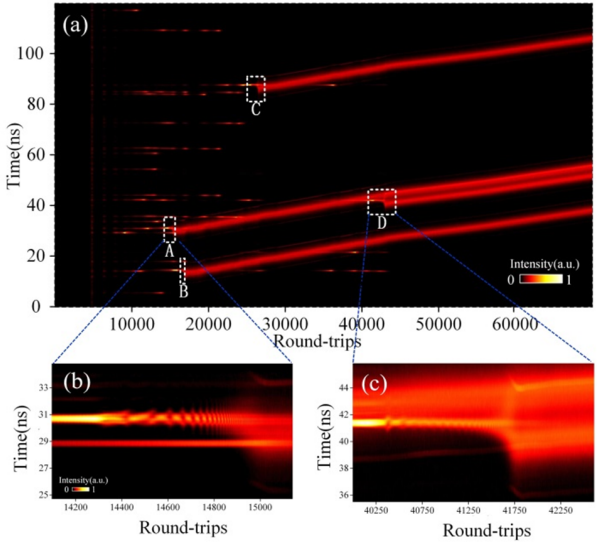

Fig. 3. (a) Real-time 2D spectral evolution dynamics of multiple soliton. (b)-(c) Beating processes of first and fourth solitons, respectively.

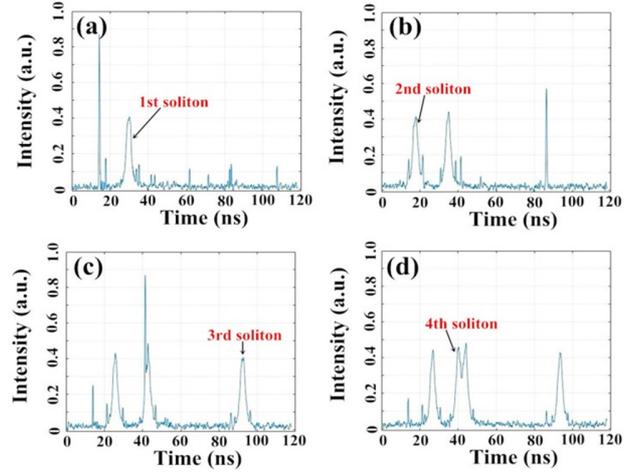

Fig. 4. Experimental real-time single-shots for the formation and evolution of multiple solitons. (a) Build-up of 1st soliton (see also Media 1). (b) Build-up of 2nd soliton (see also Media 2). (c) Build-up of 3rd soliton (see also Media 3). (d) Build-up of 4th soliton (see also Media 4).

The vanished background pulse confirms the law of jungle in laser system, which works as the feeding energy of mode-locking soliton. The difference percent of vanish background pulse also shows that the mode-locking of later soliton is become easier than that of before. The trajectory of each soliton in the operation has slightly difference between each other. For the same soliton, the trajectory is also change with respect to different round-trip. The changing points are coinciding with the time of new-born solitons. The different trajectory also indicates that the different propagated speed. According to optical Kerr effect, the group velocity $v_g$ can be given:

$$v_g = \frac{c}{n_0 + \Delta n}. \quad (1)$$

where $n_0$ is the original refractive index of fiber, $\Delta n$ is the photo-induced nonlinear refractive index of fiber, and $c$ is the light speed in vacuum. Since $\Delta n \propto I$, the differences of soliton trajectory indicate different soliton intensity and the changing point of trajectory shows the time of energy re-distribution. The angle of A-soliton trajectory at the birth of B-soliton is about 10°, while is about 8° at the birth of C-soliton and is about 6° at the birth of D-soliton. The angle of B-soliton trajectory is about 10° at first, while is about 8° at the birth of C-soliton and is about 6° at the birth of D-soliton. The angle of C-soliton trajectory is about 8° at first, while is about 6° at the birth of D-soliton. The angle of D-soliton trajectory is about 6°. The former soliton reduce its speed to coincide with the new-born soliton, which means the cavity re-distribution its energy to each soliton equally. The precise soliton intensity can be calculated by spectral integration, which gives a new way to map the energy distribution process in fiber laser. The cross sections of four round-trips with respect to the generation of each soliton are shown in Fig. 4. According to the spectral information, we can calculation the precise energy of soliton at instantaneous state. The calculated energy depicts the energy fluctuation with respect to the generation of new soliton. Figure 4 also shows the real-time output spectrum with Kelly sidebands, which originate from the constructive interference between the soliton and dispersive waves [41][42]. It can be seen that there is an exceptional high intensity CW in Fig. 4(a), which is corresponded to the second soliton in Fig. 4(b). In the same sense, the highest CW in Fig. 4(b) is corresponded to the third soliton in Fig. 4(c), while the highest CW in Fig. 4(c) is corresponded to the fourth soliton in Fig. 4(d). The four typical moments in Fig. 4 are single-shots captured from the following 4 Medias, so the entire formation and evolution of first to fourth solitons are displayed in Media 1 to 4, respectively.

In this paper, we have experimentally demonstrated the real-time observation of multi-soliton asynchronous build-up dynamics in all-PM fiber laser, which shows several unique features. Each soliton generated asynchronously. Soliton generated at different time exhibits different intensity at metastable state, and has the same intensity at stable state, indicating that the intra-cavity energy exhibits quantization property. A proportion of background pulse vanished with respect to the birth time of new soliton, which shows the law of jungle in laser system. The soliton born early has positive effect on the improvement of mode-locking environment. The detailed build-up dynamics can provide great assistant to the study of soliton generation, regulation, energy improvement, background pulse suppression, stability improvement. The observation of multi-soliton build-up dynamics may open the way to understand the physical low of ultrafast system.

**Funding.** This work was supported by the National Natural Science Foundation of China under Grants No. 61525505, No. 11774310, No. 61805212, the Key Scientific and Technological Innovation Team Project in Shaanxi Province (2015KCT-06) and the Postdoctoral Science Foundation of China under Grants No. 2017M621918.

**Acknowledgment**. The authors would like to thank Xiankun Yao and Yudong Cui for assistance and fruitful discussions.